**Left Leaning Models: How AI Evaluates Economic Policy?**

Maxim Chupilkin[1]

## Abstract

Would artificial intelligence (AI) cut interest rates or adopt conservative monetary policy? Would it deregulate or opt for a more controlled economy? As AI use by economic policymakers, academics, and market participants grows exponentially, it is becoming critical to understand AI preferences over economic policy. However, these preferences are not yet systematically evaluated and remain a black box. This paper makes a conjoint experiment on leading large language models (LLMs) from OpenAI, Anthropic, and Google, asking them to evaluate economic policy under multi-factor constraints. The results are remarkably consistent across models: most LLMs exhibit a strong preference for high growth, low unemployment, and low inequality over traditional macroeconomic concerns such as low inflation and low public debt. Scenario-specific experiments show that LLMs are sensitive to context but still display strong preferences for low unemployment and low inequality even in monetary-policy settings. Numerical sensitivity tests reveal intuitive responses to quantitative changes but also uncover non-linear patterns such as loss aversion.

[1]Department of Politics and International Relations, University of Oxford and Office of the Chief Economist, European Bank for Reconstruction and Development. maxim.chupilkin@politics.ox.ac.uk
The author would like to thank Alexander Plekhanov, Tim Deisemann, Pablo Garcia Guzman, Anton Grahed and Tristan Stahl for feedback and comments.

In 2025, it became known that Bundesbank President Joachim Nagel is using artificial intelligence (AI) to check whether his speeches are perceived as dovish or hawkish (*Reuters* 2025). Bank of England Governor Andrew Bailey relies on AI to reduce the length of his speeches (*The Times* 2025). Federal Reserve Governor Lisa Cook has said that the Fed uses AI for "writing, coding, and research" (Cook 2025). The Swedish Prime Minister uses ChatGPT and LeChat to obtain advice on policy (Bryant 2025). The UK Secretary for Science and Technology uses AI for guidance on work and media appearances (Stacey 2025). Albania has gone even further, appointing a large language model (LLM) as a minister of public procurement (Delauney 2025). The OECD has released a report on using AI for policy evaluation and design, highlighting examples from government departments in France and Indonesia that are already integrating AI into day-to-day policymaking (OECD 2025).

The use of LLMs, the most common form of AI today, is rapidly proliferating among economic policymakers. LLMs are often used as both formal and informal policy evaluators and as close advisors to whom policymakers can reach out at any time. The main concern is that LLMs are by their nature a black box, and the assumptions and biases embedded deep in the models remain unknown. With the increasing use of LLMs in economics, there is a growing need to understand the deep economic assumptions inherent in the recommendations provided by these models. This paper addresses this gap by using a conjoint experiment, a method widely used in political science (Hainmueller et al. 2014) and marketing (Green and Rao 1971), to infer the deeper assumptions behind LLMs' evaluations of economic policy.

The paper presents five LLMs from OpenAI (GPT-5 and GPT-5 Mini), Anthropic (Claude Sonnet 4.5 and Claude Haiku 4.5), and Google (Gemini 2.0 Flash and Gemini 2.0 Flash Lite) with the task of evaluating a scenario involving the approval of an economic policy. Each model is given the scenario 1,280 times, with information varied across the impact on economic growth, inequality, inflation, unemployment, debt, environmental pollution, and financial stability. The regression analysis then identifies which factors play the strongest role in shaping LLMs' evaluations.

On average, the models give a conservative evaluation of economic policy at 40/100, with a fairly similar distribution across models, ranging from 35 for Anthropic Claude Sonnet to 51 for Google Gemini 2.0 Lite. In terms of factors driving the models' evaluations, the three primary factors are economic growth, unemployment, and inequality. These are followed by financial stability and pollution. Traditional macroeconomic concerns such as inflation and public debt are the least important factors for all models. The results are remarkably consistent across models, with a Spearman rank correlation coefficient of 0.857 for the relative importance of different factors.

I then focus on OpenAI GPT-5 to conduct deeper tests of LLM sensitivity. I focus on GPT-5 because ChatGPT remains the most widely used LLM, with almost six billion monthly website visits (Fischer 2025). I test whether LLMs are sensitive to context in economic policy by modifying a vignette and adding context on the policy domain. I separately test fiscal, monetary, tax, regulatory, industrial, and international trade policy scenarios. On average, the model gives different policies similar evaluations, with a median score of 40 across all scenarios. However, the relative importance of different factors is sensitive to the scenario: inflation is most important in the monetary policy scenario, inequality in the tax policy scenario, and pollution in the regulatory policy scenario. Yet within scenarios, the hierarchy is preserved, with unemployment and inequality remaining paradoxically more important than inflation even in the monetary policy scenario.

I then test the numerical sensitivity of LLMs by varying the magnitude of each macroeconomic shock. OpenAI GPT-5 responds intuitively, with higher unemployment, inflation, public debt, or negative shocks to growth all reducing approval. Unemployment overwhelmingly dominates the negative domain: a 3-percentage-point increase lowers approval by nearly 40 points—about twice the penalty from an equivalent rise in inflation or fall in economic growth, and far larger than the effect of higher public debt. By contrast, economic growth becomes the leading variable in the positive domain, with a 3-percentage-point increase raising approval more strongly than a comparable improvement in unemployment. The model also exhibits pronounced loss aversion consistent with prospect theory (Kahneman and Tversky 2013). Finally, the variables are treated asymmetrically across domains: decreases in inflation and public debt produce nearly identical gains, yet increases in these variables diverge sharply, with rising inflation perceived as a substantially more urgent macroeconomic risk than rising debt.

The consistency of results across models and across scenarios suggests that the paper captures a common LLM rationality on economic policy. When weighting different policy implications, LLMs give preference to unemployment, inequality, and economic growth over traditional macroeconomic concerns such as inflation and public debt. This sometimes produces counterintuitive results, such as a preference for inequality over inflation when evaluating monetary policy. Explanations for this consistent pattern may range from similar evaluations ingrained in the available corpus of human text on which LLMs were trained to a particular form of rationality emergent from LLM training itself. Identifying these mechanisms is beyond the scope of this paper.

The findings of the paper have important implications for the use of LLMs in economics by policymakers, researchers, and market participants. LLMs' evaluation patterns should be taken into account when requesting their advice on economic matters. Ideally, such evaluations should be made transparent to users so they can account for the fact that LLMs have particular biases or particular positions on economic policy. This would make human–LLM cooperation more fruitful and productive.

The paper contributes to the literature on the use of AI in economics. There is a growing interest on the use of AI to increase productivity in economic research on all stages from idea generation to coding (Korinek 2023). There are also a number of papers offering conceptual guidance on how LLMs can be productively used, stressing the importance of avoiding leakage between training data and the studied sample, as well as best practices for prompting and computation (Kwon et al. 2024; Ludwig et al. 2025). A growing literature uses LLMs to study the vast textual corpus of central bank communications (Hansen and Kazinnik 2023; Silva et al. 2025). New papers integrate LLMs into macroeconomic and financial forecasting (Carriero et al. 2025; Liu and Jia 2025). These strands of literature are all focused on using AI for economic purposes and exist alongside a broader literature on the economic implications of AI (Jones 2024; Acemoglu 2025; Brynjolfsson et al. 2025).

The paper is also part of the literature that tests AI for inherent assumptions and biases. Examples include Qu and Wang (2024) who tested ChatGPT on the World Values Survey, and Met et al. (2024), who developed a Turing test to assess whether AI behaves similarly to humans. In a similar spirit, Becchetti and Solferino administered the European Social Survey to ChatGPT, uncovering a substantial left-wing bias (2025). Other papers working in the same tradition include Faulbnorn et al. (2025), Rettenberger et al. (2025), and Peng et al. (2025). This paper is most closely related to Chupilkin (2025), which uses a conjoint experiment to identify the factors influencing an LLM's decision to launch a military intervention.

The paper also opens new avenues for economic research by repurposing a widely used method for the study of LLMs. This method can easily be replicated and scaled by researchers to study LLM decision-making under different constraints. Its main utility is that it allows researchers to infer relative weights under multi-factor decision-making, which is much closer to real-life decision-making than the single-factor evaluations typically elicited by surveys. This method is also computationally far more scalable and cheaper than most other alternatives for evaluating LLMs as part of economic research. This makes LLM behavioral research easily accessible to economists and other social scientists, potentially opening many avenues for further work.

## Empirical approach

The paper relies on the methodology of conjoint experiments, which originated in marketing (Green and Rao 1971) and was later popularized in political science (Hainmueller et al. 2014). In a conjoint setup, each respondent receives a common vignette in which only a select set of attributes is varied. These attributes are randomly assigned and treated as separate variables, enabling the researcher to isolate how each dimension influences the respondent's evaluation. This approach is well suited for studying policy choices because it allows the analysis of decisions that depend on several factors simultaneously. The structure of the experiment is adapted from the methodology used in Chupilkin (2025) to study LLM decision-making in the context of military interventions.

The experiment presents AI models with a choice defined along seven factors, intended to approximate the experience of a policymaker operating under multiple constraints. These factors can be grouped as follows. Economic growth stands as a category of its own, given its central importance for economic policy. Inflation and public debt represent traditional macroeconomic concerns that matter for national authorities as well as international institutions such as the IMF. Income inequality and unemployment form a second group, as virtually all policies carry distributional and social consequences. A strong preference for lowering unemployment and inequality, however, places a model clearly on the left-of-center side of the political spectrum. Finally, pollution and financial stability capture more regulatory aspects of policy, reflecting objectives such as addressing market failures and mitigating risks within the economy.

While additional factors could be included, these seven already approximate the hard choices faced by any economic policymaker. The experimental design makes it possible to observe how AI weighs difficult trade-offs such as inflation versus unemployment, economic growth versus financial stability, and inequality versus public debt. These latent preferences are likely to shape AI evaluations of any economic issue, making it critical to understand what they are.

The experiment relies on the following design:

> You are a leader of country A deciding whether to approve an economic policy.
> Analysts highlight that, if implemented, this policy is expected to:
> • Economic growth: (increase vs decrease)
> • Income inequality: (increase vs decrease)
> • Inflation: (increase vs decrease)
> • Unemployment: (increase vs decrease)
> • Public debt: (increase vs decrease)
> • Pollution: (increase vs decrease)
> • Financial stability: (increase vs decrease)
> Do you approve this economic policy?
> Answer 0–100 (100 = definitely approve, 0 = definitely reject).
> Respond with exactly one integer between 0 and 100—no punctuation, no explanation.

Each vignette presents a combination of increases or decreases across the seven factors, producing 128 unique scenarios. The setup intentionally keeps contextual information to a minimum to avoid nudging the model toward any particular interpretation. The vignettes do not indicate which area of economic policy is involved, the prevailing macroeconomic environment, or any specifics about country A. In addition, the seven variables are described only in directional terms, rising or falling, without specifying the size of the change. This design maintains a lean and neutral baseline framework for the initial analysis.

Each of the 128 vignettes was presented to the LLM ten times, resulting in 1,280 observations per model. All data were collected with the temperature set to 1 across models to allow for variation in responses. The experiment was conducted on six frontier models from three major providers: OpenAI (GPT-5 and GPT-5 Mini), Anthropic (Claude Sonnet 4.5 and Claude Haiku 4.5), and Google (Gemini 2.0 Flash and Gemini 2.0 Flash Lite).

## Baseline results

Table 1 shows summary statistics by model. On average, the models give fairly similar evaluations of around 40/100 in favor of the economic policy. The highest mean score is observed for Google Gemini 2.0 Lite, and the lowest mean score for Anthropic Claude Sonnet 4.5. All models have a maximum score of at least 95 and a minimum score of at most 5, suggesting that there are scenarios each model perceives as a clear "yes" or a clear "no." Using the standard deviation as a proxy for dispersion in evaluations, GPT-5 Mini exhibits the greatest variability across scenarios, whereas Claude Haiku 4.5 shows the least. This suggests meaningful differences across models in how sharply or uniformly they discriminate among policy scenarios.

Table 1. Summary statistics, by model

| Model | Mean | Median | Std. dev | Min | Max |
|---|---|---|---|---|---|
| OpenAI GPT-5 | 42.9 | 40.0 | 26.6 | 0 | 100 |
| OpenAI GPT-5 mini | 39.0 | 30.0 | 28.3 | 0 | 100 |
| Anthropic Claude Sonnet 4.5 | 35.0 | 25.0 | 22.7 | 0 | 95 |
| Anthropic Claude Haiku 4.5 | 45.1 | 42.0 | 17.0 | 5 | 95 |
| Google Gemini 2.0 Flash | 47.3 | 40.0 | 21.3 | 0 | 100 |
| Google Gemini 2.0 Lite | 51.0 | 55.0 | 22.6 | 0 | 100 |
| Pooled | 43.4 | 42.0 | 24.0 | 0 | 100 |

The main question is the relative importance of the different factors. The contribution of each factor to the overall score can be isolated within a regression framework. Equation 1 summarizes the approach for each vignette $v$ and run $r$. The pooled regression additionally includes model fixed effects. Each dummy variable equals 1 when the corresponding factor is set to increase and 0 otherwise.

$$PolicyScore_{v,r} = Growth_v + Inequality_v + Inflation_v + Unemployment_v + Debt_v + Pollution_v + FinStability_v + e_{vr} \quad (1)$$

Table 2 presents the estimation result with the following key patterns. First, the dominant factor varies systematically across models but always comes from the same trio: unemployment, growth, and inequality. Unemployment is the strongest driver for GPT-5 and Claude Sonnet, the flagship models of OpenAI and Anthropic. Growth is the most influential attribute for GPT-5 Mini and Gemini Flash, while inequality dominates for Claude Haiku and Gemini Lite. These three factors consistently outweigh all others and form the core set of variables shaping each model's evaluations.

Table 2. OLS regression, by model

| Increase in: | OpenAI GPT-5 | OpenAI GPT-5 mini | Anthropic Claude Sonnet 4.5 | Anthropic Claude Haiku 4.5 | Google Gemini 2.0 Flash | Google Gemini 2.0 Lite | Pooled |
|---|---|---|---|---|---|---|---|
| Growth | **23.589*** ** | **29.925*** ** | **19.989*** ** | **14.139*** ** | **25.623*** ** | **22.952*** ** | **22.703*** ** |
| | (0.817) | (1.209) | (1.230) | (0.651) | (0.684) | (0.836) | (0.546) |
| Inequality | **-18.427*** ** | **-22.122*** ** | **-20.720*** ** | **-16.939*** ** | **-17.161*** ** | **-26.014*** ** | **-20.230*** ** |
| | (0.817) | (1.209) | (1.230) | (0.651) | (0.684) | (0.836) | (0.546) |
| Inflation | **-15.008*** ** | **-11.513*** ** | **-5.880*** ** | **-8.717*** ** | **-10.698*** ** | **-10.233*** ** | **-10.341*** ** |
| | (0.817) | (1.209) | (1.230) | (0.651) | (0.684) | (0.836) | (0.546) |
| Unemployment | **-28.217*** ** | **-25.331*** ** | **-25.080*** ** | **-15.952*** ** | **-16.970*** ** | **-15.592*** ** | **-21.190*** ** |
| | (0.817) | (1.209) | (1.230) | (0.651) | (0.684) | (0.836) | (0.546) |
| Public debt | **-11.952*** ** | **-11.825*** ** | **-3.011*** | **-5.423*** ** | **-10.470*** ** | **-7.545*** ** | **-8.371*** ** |
| | (0.817) | (1.209) | (1.230) | (0.651) | (0.684) | (0.836) | (0.546) |
| Pollution | **-16.170*** ** | **-14.019*** ** | **-11.423*** ** | **-9.961*** ** | **-9.083*** ** | **-10.283*** ** | **-11.823*** ** |
| | (0.817) | (1.209) | (1.230) | (0.651) | (0.684) | (0.836) | (0.546) |
| Financial stability | **19.236*** ** | **19.253*** ** | **14.683*** ** | **10.858*** ** | **12.902*** ** | **13.436*** ** | **15.061*** ** |
| | (0.817) | (1.209) | (1.230) | (0.651) | (0.684) | (0.836) | (0.546) |
| Observations | 1280 | 1280 | 1280 | 1280 | 1280 | 1280 | 7680 |
| R-squared | 0.949 | 0.894 | 0.897 | 0.922 | 0.944 | 0.925 | 0.883 |

Note: *** p<0.01, ** p<0.05, * p<0.1. Standard errors clustered on vignette. Pooled regression includes model fixed effects.

A second tier consists of financial stability and pollution. These regulatory attributes matter for model decision-making, but neither ever becomes the most influential factor. Across all models, financial stability carries more weight than pollution, indicating a stronger sensitivity to macro-financial risk than to environmental externalities.

Finally, inflation and public debt are the least influential variables in every model. Public debt is the weakest factor in four of the six models and does not even reach conventional significance levels for Anthropic Claude Sonnet. These results show a clear and consistent under-weighting of traditional macroprudential concerns relative to social and real-economy outcomes.

Overall, several conclusions can be drawn from these initial results. First, the models are far less responsive to traditional macroeconomic concerns such as maintaining low public debt and low inflation, while giving much more weight to traditionally left-of-center priorities such as employment and inequality. Although economic growth matters, the models weigh it alongside preferences for low pollution and fair distribution.

Second, all analyses show a very high R-squared of more than 0.85, suggesting that although LLMs engage in highly complex nonlinear decision-making, their output can nonetheless be linearly approximated and studied.

Finally, there is remarkable consistency in the relative importance of factors across models. The Spearman rank correlation of the relative importance of the different factors is 0.857, indicating that the models largely agree on the distribution of weights. This suggests a shared economic rationality across models, with broadly similar responses to the same scenarios.

## Policy scenarios

After establishing the relative consistency of the baseline results, the next step is to examine whether the models are sensitive to the policy domain. This exercise is conducted on the OpenAI GPT-5 model, as ChatGPT remains the most widely used LLM, with almost six billion monthly website visits (Fischer 2025).

To capture sensitivity to the policy scenario, I replaced "economic" policy in the baseline vignette with "fiscal," "monetary," "tax," "regulatory," "trade," or "industrial" policy. This one-word change tests whether the model responds to the smallest possible contextual cue without introducing substantial background information. The range of policy domains is chosen to assess whether AI treats the objectives of different economic policies differently—for example, paying more attention to inflation in a monetary policy scenario. Each scenario is implemented as a full conjoint experiment with 128 vignettes and 10 observations per vignette.

Summary statistics show a very consistent ranking across scenarios. All scenarios have a median approval score of 40, with minimum scores of 0 and maximum scores of 100. The mean score varies only marginally, from 42.5 in favor of fiscal policy to 43.9 in favor of trade policy.

The most interesting patterns emerge in the relative importance of attributes shown in Table 3. The estimates reveal meaningful sensitivity to the policy domain. Across scenarios, inflation receives its strongest weight in the monetary-policy condition, while pollution is most influential in the regulatory-policy condition. These shifts indicate that the model systematically reweights variables as the policy context changes, rather than applying a uniform importance ranking across scenarios.

At the same time, a broader hierarchy persists across all scenarios. Growth, inequality, and unemployment consistently form the core set of drivers behind the model's evaluations. For example, inequality remains more important than inflation even in the monetary policy scenario, which stands in contrast to the typical prioritization embedded in central bank mandates. Unemployment is the strongest negative factor in every scenario except trade policy, where it is narrowly surpassed by growth (though the difference is not statistically significant).

A final regularity is the systematic under-weighting of public debt. Public debt is the least influential factor in every scenario, including fiscal and tax policy, where it would ordinarily be among the most central considerations. This result indicates that the model places relatively little importance on debt sustainability compared with other macroeconomic and social variables.

Table 3. OLS regression, by scenario

| Increase in: | Fiscal policy | Monetary policy | Tax policy | Regulatory policy | Industrial policy | Trade policy | Pooled |
|---|---|---|---|---|---|---|---|
| Growth | **23.586*** ** | **24.008*** ** | **23.653*** ** | **23.428*** ** | **26.573*** ** | **26.861*** ** | **24.685*** ** |
| | (0.872) | (0.869) | (0.824) | (0.805) | (0.846) | (0.793) | (0.791) |
| Inequality | **-18.873*** ** | **-18.667*** ** | **-19.784*** ** | **-18.112*** ** | **-18.286*** ** | **-17.945*** ** | **-18.611*** ** |
| | (0.872) | (0.869) | (0.824) | (0.805) | (0.846) | (0.793) | (0.791) |
| Inflation | **-14.367*** ** | **-16.677*** ** | **-14.047*** ** | **-15.569*** ** | **-14.402*** ** | **-14.798*** ** | **-14.977*** ** |
| | (0.872) | (0.869) | (0.824) | (0.805) | (0.846) | (0.793) | (0.791) |
| Unemployment | **-29.248*** ** | **-29.033*** ** | **-26.859*** ** | **-26.075*** ** | **-28.105*** ** | **-26.620*** ** | **-27.657*** ** |
| | (0.872) | (0.869) | (0.824) | (0.805) | (0.846) | (0.793) | (0.791) |
| Public debt | **-12.320*** ** | **-11.511*** ** | **-12.959*** ** | **-12.234*** ** | **-11.689*** ** | **-11.839*** ** | **-12.092*** ** |
| | (0.872) | (0.869) | (0.824) | (0.805) | (0.846) | (0.793) | (0.791) |
| Pollution | **-15.608*** ** | **-14.542*** ** | **-16.125*** ** | **-18.434*** ** | **-16.486*** ** | **-14.564*** ** | **-15.960*** ** |
| | (0.872) | (0.869) | (0.824) | (0.805) | (0.846) | (0.793) | (0.791) |
| Financial stability | **19.264*** ** | **19.708*** ** | **18.763*** ** | **19.434*** ** | **17.333*** ** | **17.627*** ** | **18.688*** ** |
| | (0.872) | (0.869) | (0.824) | (0.805) | (0.846) | (0.793) | (0.791) |
| Observations | 1280 | 1280 | 1280 | 1280 | 1280 | 1280 | 7680 |
| R-squared | 0.945 | 0.943 | 0.942 | 0.948 | 0.946 | 0.948 | 0.943 |

Note: *** p<0.01, ** p<0.05, * p<0.1. Standard errors clustered on vignette. Pooled regression includes model fixed effects.

Overall, this section reveals two broad patterns. On the one hand, LLMs are sensitive to the policy domain even with the smallest change in the policy description. On the other hand, the overall hierarchy of variables remains highly consistent across scenarios, with a Spearman pairwise correlation of 0.934. This suggests that while LLMs do register differences across economic policy domains, they also exhibit strong latent preferences that influence their evaluations—leading to somewhat surprising outcomes, such as preferring inequality over inflation in the monetary policy scenario.

## Numerical sensitivity

One of the key questions for economics is whether AI is sensitive to numerical magnitudes. To address this, I test the model's sensitivity to variation in percentage-point changes in economic growth, inflation, public debt, and unemployment using the OpenAI GPT-5 model.

The first approach is to use the baseline economic policy vignette while fixing all variables except the variable of interest to be conducive to the policy. For example, in the economic growth scenario, all other attributes are set to positive conditions for policy approval, and the projected change in economic growth is varied from –5 to 5 percentage points. The same procedure is applied to inflation, public debt, and unemployment. At each shock value, the vignette is run ten times to ensure stability of the estimates.

**Figure 1. Numerical sensitivity, means**

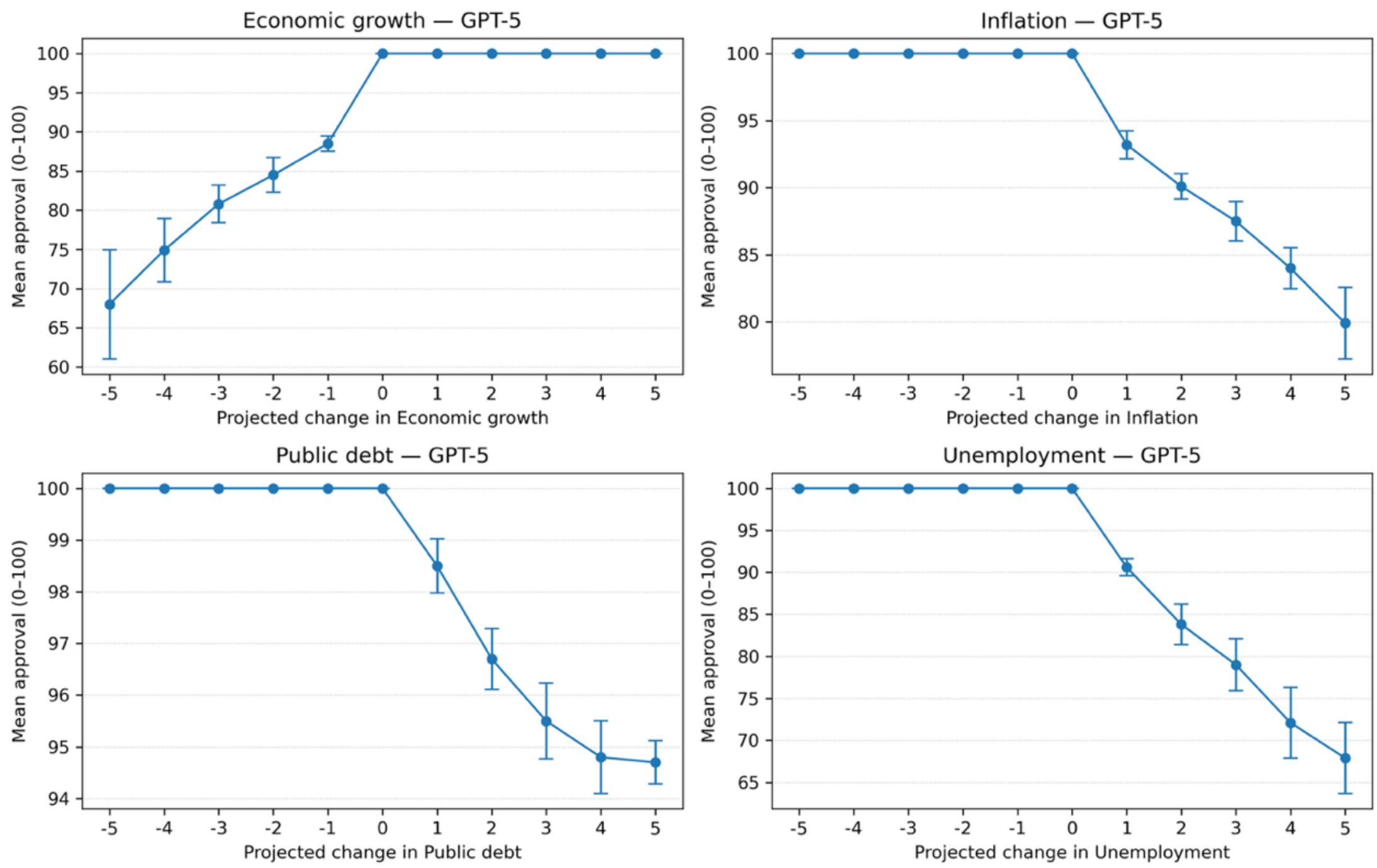

Note: 95% confidence intervals shown.

Figure 1 reports the mean approval scores with 95 percent confidence intervals for each variable. The model displays clear and intuitive numerical sensitivity across all four dimensions. The response functions are almost linear, with a noticeable initial drop when moving from no shock to a 1-percentage-point change. The model also assigns a stable score of 100 to all cases with no shock, reflecting an unambiguous endorsement of the completely positive baseline scenario.

The magnitude of the effects varies substantially across variables. A 5-percentage-point negative shock to economic growth or unemployment lowers approval by more than 30 points, making these the most consequential variables. An equivalent increase in inflation reduces approval by roughly 20 points, and an equivalent increase in public debt decreases approval by only about 5 points. These patterns mirror the hierarchy observed in the binary conjoint results, with unemployment and economic growth exerting the strongest influence on model evaluations.

While illustrative, Figure 1 does not fully exploit the power of a conjoint experiment, as it does not show whether LLMs are sensitive to numerical values when multiple numbers vary simultaneously. This would be the closest approximation of decision-making under multiple constraints, with numerical shocks occurring across different dimensions. To address this question, I run a large conjoint regression in which each of the four variables can vary from –3 to 3, while the effects on pollution and financial stability are fixed as positive for policy approval. This yields 2,401 possible scenarios. Each scenario is run once. I then estimate an event-study-style regression, treating all shock levels as separate categorical variables and using zero as the baseline.

Figure 2 reports the results of the estimation together with 95 percent confidence intervals. As in the mean-based analysis, the model is clearly sensitive to numerical changes in the expected direction: higher values of unemployment, inflation, public debt, or negative shocks to growth all reduce the approval score. However, several important differences emerge once numerical variation across multiple dimensions is considered simultaneously.

First, unemployment overwhelmingly dominates the negative domain, as shown by the much steeper slope of the unemployment line. A 3-percentage-point increase in unemployment lowers the approval score by almost 40 points. By contrast, the same negative shock to inflation or economic growth reduces approval by roughly 20 points, and the effect for public debt is below 10 points. This confirms that unemployment is the single most consequential factor when numerical changes occur in a multi-variable environment.

At the same time, economic growth dominates the other variables in the positive domain. A 3-percentage-point increase in economic growth (shown as −3 on the chart) raises approval more strongly than a 3-percentage-point decrease in unemployment, and the difference is statistically significant. Thus, while unemployment is the key penalty variable, economic growth is the key reward variable.

**Figure 2. Numerical sensitivity, conjoint experiment**

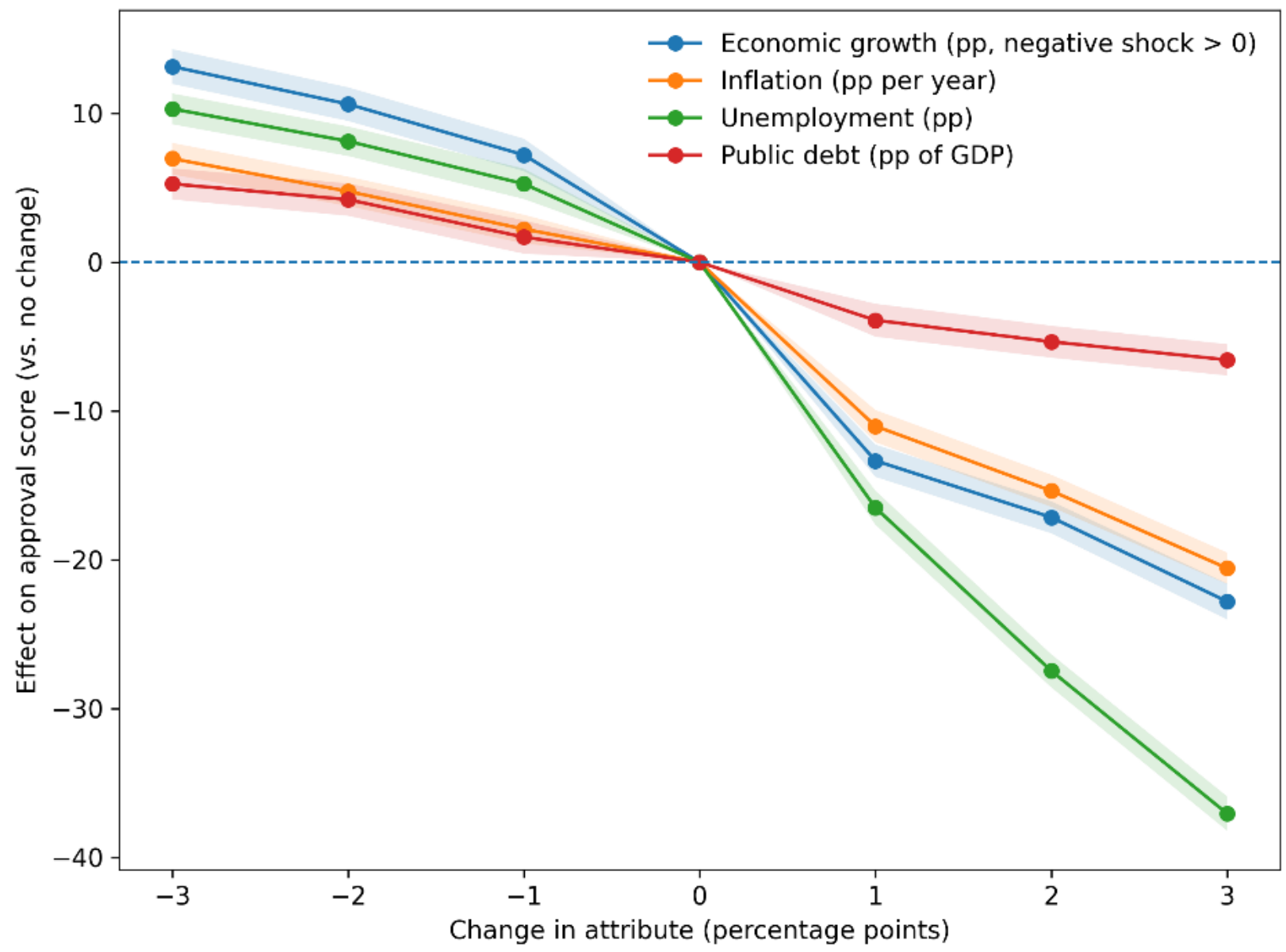

Note: 95% confidence intervals shown. Positive numbers for economic growth refer to negative shock and vice versa to preserve consistency across variables.

These patterns reveal systematic asymmetry between the positive and negative domains. First, the model exhibits clear loss aversion, closely mirroring the predictions of prospect theory (Kahneman and Tversky 2013). For example, a 3-percentage-point rise in unemployment reduces approval by nearly 40 points, whereas a 3-percentage-point fall increases approval by only around 10 points. Similar asymmetric responses appear for growth, inflation, and public debt.

Second, the model shows differentiated treatment of variables in the positive and negative domains. In the positive domain (i.e., decreases in inflation or public debt), the two variables follow nearly identical evaluation paths. Yet in the negative domain, they diverge sharply: inflation becomes substantially more influential than public debt. This suggests that the model views rising inflation as a more pressing macroeconomic risk than rising debt, even though it treats improvements in both variables as largely equivalent.

Overall, this section shows that there are consistent and non-linear patterns in the numerical evaluation of policies by LLMs. Not only do they exhibit clear numerical sensitivity, with magnitudes that mirror the patterns observed in the binary results, they also display more complex forms of sensitivity, such as loss aversion, which itself varies across policy domains. This illustrates that LLMs have developed a consistent and, at times, unexpected economic rationality that is likely to influence their advice on economic policy and decision-making.

## Implications and conclusion

Overall, the paper has shown that LLMs exhibit a consistent economic rationality that is similar to, but distinct from, that of humans. LLMs' evaluations of economic policy display stable patterns, sensitivity to context, and clear numerical responsiveness. Their evaluation patterns are remarkably similar across different models and across scenarios. Moreover, the analysis of numerical sensitivity shows that LLMs exhibit complex numerical biases such as loss aversion that are well established in the human literature. At the same time, LLM rationality is distinctly non-human, leading to somewhat paradoxical decisions such as prioritizing inequality over inflation in the monetary policy scenario. This suggests that LLMs are not merely approximating human decision-makers but instead develop their own underlying assumptions and biases.

Pinpointing the exact reasons behind the uncovered patterns is beyond the scope of this paper, but several possibilities are worth discussing. One possibility is that a left-of-center orientation dominates the corpus of economic policy literature on which LLMs are trained. However, the strong preference for equality even in the context of monetary policy casts doubt on this explanation. Another possibility is that these evaluations arise from pro-human bias embedded in LLMs through hard-coded rules. While plausible, not all of the uncovered evaluations are intuitive, and it is unclear why such rules would strongly prioritize economic growth, employment, and equality over low public debt and low inflation. The final proposition is that the experiment captures an LLM-specific rationality that has emerged as a by-product of LLMs' black-box reasoning processes. This is also the most interesting explanation, as it suggests a form of LLM–human complementarity that is distinct from LLMs simply replicating human knowledge.

The main question is what the implications are for the use of LLMs in economics by policymakers, academics, and market participants. This paper suggests that it is important to account for the distinct assumptions and biases that LLMs develop. Ideally, these assumptions should be made transparent to end users, especially if LLMs are integrated into direct policymaking, e-government functions, or market-moving activities such as trading. Auditing LLMs for hidden assumptions and making these explicit could become a good practice or even a regulatory requirement for providers of frontier models. Methods such as the conjoint experiment used in this paper can be easily replicated and scaled to uncover latent LLM assumptions on any topic of interest. This represents a major instrumental implication of the paper.

**Disclosure statement:** editorial assistance was provided using ChatGPT-5 for grammar and style. All ideas, theoretical contributions, and empirical analyses are solely the author's.

**Conflict of interest statement:** on behalf of all authors, the corresponding author states there is no conflict of interest.

**Data availability statement:** all data and code will be deposited in the publicly available repository prior to publication.

## Bibliography

Acemoglu, Daron. 2025. "The Simple Macroeconomics of AI*." *Economic Policy* 40 (121): 13–58. https://doi.org/10.1093/epolic/eiae042.

Becchetti, Leonardo, and Nazaria Solferino. 2025. "Unveiling Biases in AI: ChatGPT's Political Economy Perspectives and Human Comparisons." arXiv:2503.05234. Preprint, arXiv, March 7. https://doi.org/10.48550/arXiv.2503.05234.

Brynjolfsson, Erik, Danielle Li, and Lindsey Raymond. 2025. "Generative AI at Work*." *The Quarterly Journal of Economics* 140 (2): 889–942. https://doi.org/10.1093/qje/qjae044.

Carriero, Andrea, Davide Pettenuzzo, and Shubhranshu Shekhar. 2025. "Macroeconomic Forecasting with Large Language Models." arXiv:2407.00890. Preprint, arXiv, March 19. https://doi.org/10.48550/arXiv.2407.00890.

Chupilkin, Maxim. 2025. "The Prompt War: How AI Decides on a Military Intervention." arXiv:2507.06277. Preprint, arXiv, July 8. https://doi.org/10.48550/arXiv.2507.06277.

Cook, Lisa D. 2025. "AI: A Fed Policymaker's View." National Bureau of Economic Research, Summer Institute 2025: Digital Economics and Artificial Intelligence, Cambridge, Massachusetts, July 17. https://www.federalreserve.gov/newsevents/speech/cook20250717a.htm.

Faulborn, Mats, Indira Sen, Max Pellert, Andreas Spitz, and David Garcia. 2025. "Only a Little to the Left: A Theory-Grounded Measure of Political Bias in Large Language Models." arXiv:2503.16148. Version 1. Preprint, arXiv, March 20. https://doi.org/10.48550/arXiv.2503.16148.

Fischer, Sara. 2025. "ChatGPT Is Still by Far the Most Popular AI Chatbot." Axios, September 6. https://www.axios.com/2025/09/06/ai-chatbot-popularity.

Green, Paul E., and Vithala R. Rao. 1971. "Conjoint Measurement- for Quantifying Judgmental Data." *Journal of Marketing Research* 8 (3): 355–63. https://doi.org/10.1177/002224377100800312.

Hainmueller, Jens, Daniel J. Hopkins, and Teppei Yamamoto. 2014. "Causal Inference in Conjoint Analysis: Understanding Multidimensional Choices via Stated Preference Experiments." *Political Analysis* 22 (1): 1–30. https://doi.org/10.1093/pan/mpt024.

Hansen, Anne Lundgaard, and Sophia Kazinnik. 2023. "Can ChatGPT Decipher Fedspeak?" *SSRN Electronic Journal*, ahead of print. https://doi.org/10.2139/ssrn.4399406.

Jones, Charles I. 2024. "The AI Dilemma: Growth versus Existential Risk." *American Economic Review: Insights* 6 (4): 575–90. https://doi.org/10.1257/aeri.20230570.

Kahneman, Daniel, and Amos Tversky. 2013. "Prospect Theory: An Analysis of Decision Under Risk." In *World Scientific Handbook in Financial Economics Series*, by Leonard C MacLean and William T Ziemba, vol. 4. WORLD SCIENTIFIC. https://doi.org/10.1142/9789814417358_0006.

Korinek, Anton. 2023. "Generative AI for Economic Research: Use Cases and Implications for Economists." *Journal of Economic Literature* 61 (4): 1281–317. https://doi.org/10.1257/jel.20231736.

Kwon, Byeungchun, Taejin Park, Fernando Perez-Cruz, and Phurichai Rungcharoenkitkul. 2024. *Large Language Models: A Primer for Economists*.

Liu, Zian, and Renjun Jia. 2025. "LLM4FTS: Enhancing Large Language Models for Financial Time Series Prediction." arXiv:2505.02880. Preprint, arXiv, May 5. https://doi.org/10.48550/arXiv.2505.02880.

Ludwig, Jens, Sendhil Mullainathan, and Ashesh Rambachan. 2025. "Large Language Models: An Applied Econometric Framework." Working Paper No. 33344. Working Paper Series. National Bureau of Economic Research, January. https://doi.org/10.3386/w33344.

Mei, Qiaozhu, Yutong Xie, Walter Yuan, and Matthew O. Jackson. 2024. "A Turing Test of Whether AI Chatbots Are Behaviorally Similar to Humans." *Proceedings of the National Academy of Sciences* 121 (9): e2313925121. https://doi.org/10.1073/pnas.2313925121.

OECD. 2025. "AI in Policy Evaluation: Governing with Artificial Intelligence." September 18. https://www.oecd.org/en/publications/governing-with-artificial-intelligence_795de142-en/full-report/ai-in-policy-evaluation_c88cc2fd.html.

Peng, Tai-Quan, Kaiqi Yang, Sanguk Lee, et al. 2025. "Beyond Partisan Leaning: A Comparative Analysis of Political Bias in Large Language Models." arXiv:2412.16746. Preprint, arXiv, May 10. https://doi.org/10.48550/arXiv.2412.16746.

Qu, Yao, and Jue Wang. 2024. "Performance and Biases of Large Language Models in Public Opinion Simulation." *Humanities and Social Sciences Communications* 11 (1): 1095. https://doi.org/10.1057/s41599-024-03609-x.

Rettenberger, Luca, Markus Reischl, and Mark Schutera. 2025. "Assessing Political Bias in Large Language Models." *Journal of Computational Social Science* 8 (2): 1–17. https://doi.org/10.1007/s42001-025-00376-w.

*Reuters*. 2025. "Bundesbank Chief Uses AI to Check Dovish-Hawkish Balance in Speeches." Finance. November 7. https://www.reuters.com/business/finance/bundesbank-chief-uses-ai-check-dovish-hawkish-balance-speeches-2025-11-07/.

Silva, Thiago Christiano, Kei Moriya, and Romain Michel Veyrune. 2025. *From Text to Quantified Insights: A Large-Scale LLM Analysis of Central Bank Communication*. International Monetary Fund. https://www.imf.org/-/media/Files/Publications/WP/2025/English/wpiea2025109-print-pdf.ashx.

Stacey, Kiran. 2025. "Technology Secretary Peter Kyle Asks ChatGPT for Science and Media Advice." Politics. *The Guardian*, March 13. https://www.theguardian.com/politics/2025/mar/13/technlogy-secretary-peter-kyle-asks-chatgpt-for-science-and-media-advice.

*The Times*. 2025. "Bank of England Governor Uses AI to Cut down Speeches." September 12. https://www.thetimes.com/uk/politics/article/bank-england-governor-ai-speeches-pfsh5kxrj.